\documentclass[journal,twocolumn]{IEEEtran}
\usepackage{amssymb}
\usepackage{amsfonts}
\usepackage{amsmath}
\usepackage{algorithm}
\usepackage{algorithmic}

\usepackage{graphicx,subfigure,amsmath,amssymb,cite}

\usepackage[dvips]{color}
\usepackage{float}
\ifCLASSINFOpdf
\else
\fi
\begin{document}

\title{UAV Deployment Optimization for Secure Precise Wireless Transmission}

\author{Tong Shen, Guiyang Xia, Jingjing Ye, Lichuan Gu, Xiaobo Zhou and Feng Shu
\thanks{Tong Shen Guiyang Xia, Xiaobo Zhou, and Lichuan Gu are the School of Information and Computer, Anhui Agricultural University, 230036, CHINA. (Email: shentong0107@163.com, xiaguiyang@njust.edu.cn, zxb@fynu.edu.cn, glc@ahau.edu.cn). }
\thanks{Feng Shu is with School of Information and Communication Engineering, HaiNan University, HaiNan, 570228, China(E-mail: shufeng0101@163.com).}
\thanks{Jingjing Ye is with Mingguang Meteorological Mureau, Anhui, 239400, China.(E-mail: 23139008@qq.com).}
}
\maketitle

\begin{abstract}
This paper develops an unmanned aerial vehicle (UAV) deployment scheme in the context of the directional modulation-based secure precise wireless transmissions (SPWTs), where the optimal UAV position for the SPWT is derived to maximum the secrecy rate (SR) without injecting any artificial noise (AN) signaling. To be specific, the proposed scheme reveals that the optimal position of UAV for maximizing the SR performance has to be placed at the null space of Eves channel, which impels the received energy of the confidential message at the unintended receiver deteriorating to zero whilst benefits the one at the intended receiver achieving its maximum value. Finally, simulation results verify the optimality of our proposed scheme in terms of the achievable SR performance.
\end{abstract}

\section{Introduction}
As a potential candidate for benefiting wireless transmission in physical layer \cite{Wu2018JSAC,ChenXM2017,WangHM2022,YanSH2022}, directional modulation (DM) technique has attracted extensive attention resulted from that DM can transmit confidential signal to a specific direction \cite{Daly2009Directional,Yuan2015,Hu2016DM}. Technically, DM is generally employed in two fashions: one is implemented on radio frequency (RF) frontend employing the phased array (PA) to optimize the phase shift \cite{Daly2009Directional} while another one is implemented on the baseband by utilizing orthogonal vector \cite{Yuan2015} or beamforming operations \cite{Hu2016DM}. Due to the broadcast nature of DM-based wireless communications, security issue is unavoidable since the distance dimension of DM is threatened by the unintended facility, although DM has an ability of enabling the security in direction aspect. To address this problem, the authors of \cite{LFDA} proposed a linear frequency diverse array (LFDA) method to achieve SPWT. However, the direction angle and distance achieved by LFDA may be coupled, which means that there may exist multiple directions and distances receiving the same confidential message as the desired users received. As a further advance, a number of DM schemes taking the advantage of random frequency diverse array (RFDA) \cite{Nusenu2022,WangWQ2022TWC,WangWQ2022TSP} have been proposed to achieve secure precise wireless transmission (SPWT) for both the direction and distance domains in which the frequency is randomly allocated on each transmit antenna and the correlation direction and distance are decoupled.. 

Thereafter, SPWT develops repidly. In \cite{ShenTVT}, the authors proposed two practical random subcarrier selection schemes which increase the randomness of subcarriers based on the random FDA, moreover it improves the stability of the SPWT system since a more random selected subcarriers lead to a better performance. In \cite{ShenWCL}, a hybrid beamforming scheme with hybrid digital and analog SPWT structure was proposed, which reduces the circuit budget with low computational complexity and comparable secrecy performance, it significantly increases the practicability of SPWT system. Since the wireless propagation channel results in the security problem becoming more formidable due to the accessibility for diverse devices. For \cite{Shu2018SPWT} as example, the authors clarify that, the energy of the main lobe is always formed around the desired receiver, but a number of the non-negligible side lobe remain having comparatively appreciable power. It is indicated that the position of the transmitter affects the distribution of those lobes, and then leading to a security risk. To elaborate a little, when the eavesdropper is located on the side-lobe peak, the achievable secrecy rate (SR) performance will be gravely degraded. Therefore, deploying the transmitter at an appropriate position has an important meaning for the secrecy performance.

As a matter of fact, the most existing works regarding the SPWT focus on static scenarios, which severely limits its application scenarios. Considering that unmanned aerial vehicle (UAV) has been widely utilized in wireless communications due to bringing extensive benefits (e.g., high probability air-to-ground channel \cite{Zhou2019} and mobility controllable\cite{Wu2018}). Moreover, UAV transmission technology has been researched wildly. In \cite{ZhouUAV2021}, the authours consider UAV networks for collecting data securely and covertly from ground users, a full-duplex UAV intends to gather confidential information from a desired user through wireless communication and generate artificial noise (AN) with random transmit power in order to ensure a negligible probability of the desired user's transmission being detected by the undesired users. The authours in \cite{HuUAV2020} proposed a detection strategy based on multiple antennas with beam sweeping to detect the potential transmission of UAV in wireless networks. In \cite{LiUAV2021}, a novel framework is established by jointly utilizing multiple measurements of received signal strength from multiple base stations and multiple points on trajectory to improve the localization precision of UAV. In \cite{ChengUAV2023}, the authors considered the region constraint and proposed a received-signal-strength-based optimal scheme for drones swarm passive location measurement                                                                                                                                                                                                                                                                                            . Thus, this work we consider SPWT in the context of UAV networks, which not only extends the static scenarios to the dynamic situations but also matches the stringent requirement of SPWT for line-of-sight (LoS) communication link from the transmitter to receiver.

Note that in the previous works (e.g., \cite{Nusenu2022,WangWQ2022TWC,WangWQ2022TSP}), the FDA technique is generally employed to determine a specific position, while this work casts off the high-cost FDA scheme but ingeniously achieves this goal by fully taking advantage of the angle information in the three-dimensional (3D) space. Against this background, this letter considers an SPWT scheme with the aid of a UAV to improve the system’s security level. For a removable UAV transmitter, an analytical solution to the optimal UAV position is derived for reducing the computational complexity. Finally, simulation results show the efficiency of the proposed UAV deployment scheme in terms of the achievable SR performance.

\indent The remainder of this paper is organized as follows:In Section II,our system model of proposed UAV SPWT is described. Then the secrecy capacity performance based on UAV SPWT structure and proposed UAV deployment scheme is analyzed in Section III. Section IV presents the simulation and the analysis. Finally, the conclusion is drawn in Section V.\\
\indent Notations: In this paper,scalar variables are denoted by italic symbols,vectors and matrices are denoted by letters of bold upper case, bold lower case, respectively. Sign ${\left(  \cdot  \right)^T}$, ${\left(  \cdot  \right)^*}$ $tr\left(  \cdot  \right)$ and ${\left(  \cdot  \right)^H}$ denote transpose,conjugation,trace and conjugate transpose respectively. $\left\|  \cdot  \right\|$ and $\left|  \cdot  \right|$ denote the norm and modulus respectively. ${\mathbf{I}_{M \times N}}$ denotes the $M \times N$ identity matrix and $\mathbb{E}\left[  \cdot  \right]$ denotes expectation operation

%

\section{System Model}
\begin{figure}[t]
\centering
\includegraphics[width=0.4\textwidth]{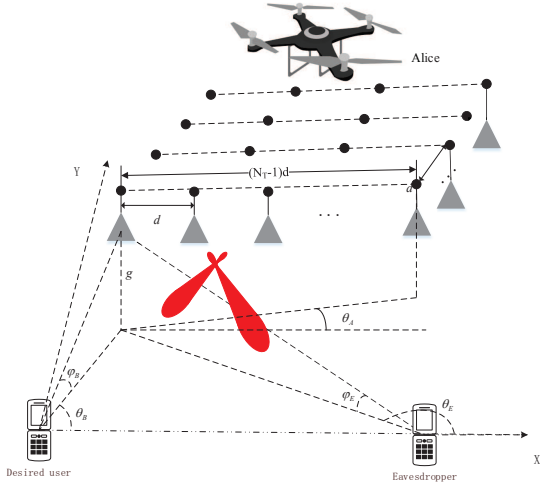}\\
\caption{System model.}\label{system}
\end{figure}

As shown in Fig. \ref{system}, our considered SPWT system is composed of an UAV, a desired user (Bob) and an eavesdropper (Eve). Herein, both Bob and Eve are equipped with a single antenna while the UAV is equipped with a $M\times N$ rectangular antenna array, namely the distance between any two adjacent antenna elements is identical.
For expression convenience, we set Bob as the origin and the ray formed from Bob to Eve is defined as the positive direction of the X-axis.
Moreover, we considered that Bob and Eve are ground users, i.e. the Z-axis coordinates of both Bob and Eve are $0$. We assume that the UAV flies at a predetermined height $g$ and parallel to the ground. The channels between the users (Bob or Eve) are assumed as LoS channels which have been widely used in UAV communication scenarios \cite{ZhouUAV2019}. What is more important, it is difficult for SPWT of applying in None-LoS (NLoS) channels, the reasons are as follows: Firstly, as NLOS channel is independent on $\theta$ and $R$, the designed beamforming vectors have the possibility of transmitting the confidential signal to any location. As a result, the security performance might be seriously degraded. Secondly, as for NLOS channels, it changes with time, the designed beamforming vectors can be only applied for a specific time. Therefore, with the designed beamforming vectors in NLOS channel, confidential message may be transmitted to any location as time goes. Lastly, in our proposed scheme, the invorked artificial noise (AN) has an ability of disturbing the signal received at Eve, but having a negligible effect on Bob. However, in NLOS channels, there exists an effect of gathering AN for Bob, thus resulting in a serious secrecy rate performance degradation at Bob. Thus, the assumption of the LoS channel in our proposed UAV and SPWT system is completely reasonable.

Due to the UAV serving the ground users, the relative positions of Bob and Eve is assumed to be known by the UAV. Rationality analysis of this assumption is as follows. For Bob, it is reasonable that the UAV knows his target user's location. For Eve, we consider a large number of users in our system, and all the users' location (including desired users, undesired users and eavesdroppers) are known for Alice. The part of undesired users are legitimate users in other time periods and they will not eavesdrop the the confidential message. The other part of eavesdroppers are illegitimate users in all the time, however, the eavesdroppers hidden in all the users, who is an undesired user or who is an eavesdropper can not be determined. Thus, we can not determine which user is an eavesdropper, but can consider a user who is most likely to be a eavesdropper. This scheme is obviously more reasonable and practical, and thus the Eve in our proposed system model is the most likely eavesdropper, and we try to prevent his eavesdropping by our proposed scheme. In conclusion, the assumption that the relative positions of Bob and Eve is known to the UAV.

According to the definition, the 3D coordinates of Bob is $(0,0,0)$, of Eve is $(X_E,0,0)$ while of the UAV is $(X_A,Y_A,g)$. Moreover, $\theta_A\in (0,2\pi)$ is the yawing angle between UAV-flied direction and X-axis.
Then, based on the established 3D rectangular coordinate system, the angle relationship among UAV, Bob and Eve can be derived. Specifically, the azimuth angle of Bob with respect to UAV follows $\sin\theta_B=\frac{Y_A}{\sqrt{X_A^2+Y_A^2}}$ and $\cos\theta_B=\frac{X_A}{\sqrt{X_A^2+Y_A^2}}$, while the azimuth angles of Eve with respect to UAV satisfies $\sin\theta_E=\frac{Y_A}{\sqrt{(X_A-X_E)^2+(Y_A)^2}}$ and  $\cos\theta_E=\frac{X_A-X_E}{\sqrt{(X_A-X_E)^2+(Y_A)^2}}$. Owing to the existence of $\theta_A$, the azimuth angles with respect to the antenna array of the UAV are given by $\theta^{'}_B=\theta_B-\theta_A$ and $\theta^{'}_E=\theta_E-\theta_A$, respectively.
Similarly, the pitch angles of Bob and of Eve related to the UAV can be respectively derived as $\sin\varphi_B=\frac{g}{\sqrt{(X_A)^2+(Y_A)^2+g^2}}$ and $\sin\varphi_E=\frac{g}{\sqrt{(X_A-X_E)^2+(Y_A)^2+g^2}}$, where both $\varphi_B$ and $\varphi_E$ satisfy $\varphi_B, \varphi_E \in \left[0, {\pi}/{2}\right]$.

Due to UAV being a $M\times N$ rectangular antenna array, we define the array parallel to the flight direction as the row, whilst arrays perpendicular to the direction of flight as columns. As such, the subscripts $m$ and $n$ are used to denote the $m$-th row and $n$-th column of the rectangular antenna array.
Then, the steering vector for the antenna array is expressed as
\begin{align}\label{h}
\mathbf{h}(\theta,\varphi)=\frac{1}{\sqrt{MN}}[e^{j2\pi \psi_{1,1}}\ldots e^{j2\pi \psi_{m,n}}\ldots e^{j2\pi \psi_{M,N}}],
\end{align}
where $\theta$ and $\varphi$ respectively denote the receiver's azimuth angle and the pitch angle relative to the UAV, where $\psi_{m,n}$ is
\begin{align}\label{psimn}
\psi_{m,n}=-\frac{f_c}{c}[(m-1)d\cos \theta^{'} +(n-1)d\sin \theta^{'} ]\cos \varphi.
\end{align}
Herein, $f_c$ is the central carrier frequency, $d=c/(2f_c)$ is the element spacing in the transmit antenna array and $c$ is the speed of light. Substituting $(\theta^{'}_B,\varphi_B)$ and $(\theta^{'}_E,\varphi_E)$ into (\ref{h}) and (\ref{psimn}), respectively, we obtain the steering vector $\mathbf{h}(\theta^{'}_B,\varphi_B)$ and $\mathbf{h}(\theta^{'}_E,\varphi_E)$.

At the baseband, the transmit signal can be expressed as
\begin{align}\label{s}
s=\sqrt{\alpha P_s}\mathbf{v}x+\sqrt{(1-\alpha) P_s}\mathbf{w},
\end{align}
where $\alpha$, $P_s$ and $x$ refer to the power allocation factor, total transmit power and a complex symbol following $\mathbb{E}[|x|^2]=1$. Besides, $\mathbf{v}\in \mathbb{C}^{MN \times 1}$ and $\mathbf{w}\in \mathbb{C}^{MN \times 1}$ denote the beamforming and AN vectors, respectively.

To achieve precise transmission, $\mathbf{v}=\mathbf{h}(\theta_B,\varphi_B)$ is set to maximize the received power of the confidential signal at Bob, while $\mathbf{w}=[\mathbf{I}_{M N}-\mathbf{h}(\theta^{'}_B,\varphi_B)\mathbf{h}^H(\theta^{'}_B,\varphi_B)]\mathbf{z}$
is to project the AN into the null space of Bob, where $\mathbf{z}$ is AN vector consisted of $MN$ complex Gaussian variables with normalized power, i.e., $\mathbf{z}\sim \mathcal{CN}(0,\mathbf{I}_{M N})$.
Notably, UAV is usually an aerial platform serving for terrestrial nodes, hence all the communication channels follow light of sight (LOS) model. Then, the received signal at Bob can be expressed as
\begin{align}\label{yb}
y_B=\sqrt{\alpha P_s}\mathbf{h}_B^H\mathbf{v}x+\sqrt{(1-\alpha) P_s}\mathbf{h}_B^H\mathbf{w}+n_B = \sqrt{\alpha P_s}x+n_B,
\end{align}
while the received signal at Eve is
\begin{align}\label{ye}
y_E=\sqrt{\alpha P_s}\mathbf{h}_E^H\mathbf{v}x+\sqrt{(1-\alpha) P_s}\mathbf{h}_E^H\mathbf{w}+n_E.
\end{align}
In (\ref{yb}) and (\ref{ye}), $\mathbf{h}_B$ and $\mathbf{h}_E$ are the abbreviations of $\mathbf{h}(\theta^{'}_B,\varphi_B)$ and $\mathbf{h}(\theta^{'}_E,\varphi_E)$, respectively. Moreover, $n_B$ and $n_E$ are the additive white Gaussian noises (AWGNs) at Bob and Eve satisfying $n_B\sim \mathcal{CN} (0,\sigma^2_B)$ and $n_E\sim \mathcal{CN} (0,\sigma^2_E)$.

\section{Proposed UAV position selection scheme}

In accordance with (\ref{yb}), the SPWT employs $\textbf{v}$ to impel the power of the confidential message at the desired receiver achieving its maximum. However, such a conventional SPWT does not guarantee the power received by Eve at a minimum level, since the channel spanning from UAV to Eve (i.e., $\mathbf{h}_E$) is affect by the relative position of them.
As a result, the security for the transmission of the confidential signal is unable to completely ensured. To further enhance the security level of the UAV-dominated moving network, we maximize the security rate (SR) performance by deploying the maneuverable UAV for shifting the channel attribute.

Traditionally, the SR performance can be characterized by
$\mathrm{SR}=\log(1+\mathrm{SINR_B})-\log(1+\mathrm{SINR_E})$, where $\mathrm{SINR_B}$ and $\mathrm{SINR_E}$ refer to the signal-to-interference-and-noise ratio of Bob and of Eve, respectively.
In SPWT networks, $\mathbf{v}$ has been assigned as $\mathbf{h}_B$ for benefiting $\mathrm{SINR}_B$ to arrive its maximum value. Naturally, the optimization problem related to $\theta^{'}_E$ and $\varphi_E$ can be simplified as
\begin{align} \nonumber
\min \limits_{\theta^{'}_E,\varphi_E} ~\mathrm{SINR}_E=\frac{\alpha P_s|\mathbf{h}^H_E \mathbf{h}_B|^2}{(1-\alpha) P_s\mathbf{h}^H_E \mathbf{w}+\sigma^2_E} \\ \label{opt1}
s.t.~~ 0 \leq \varphi_E\leq \frac{\pi}{2},~ 0 \leq \theta^{'}_E\leq 2\pi.
\end{align}
Considering the fact that $\mathrm{SINR}_E$ is certainly nonnegative, we associate that $\mathrm{SINR}_E$ arrives its minimum when the numerator of the objective function of \eqref{opt1} is equal to $0$. Pertinently, we expand the term $\mathbf{h}^H_E \mathbf{h}_B$ as
\begin{align}\label{hehb}
\mathbf{h}^H_E \mathbf{h}_B=&\frac{1}{MN}\sum \limits_{m,n}^{M,N} e^{j\pi [(m-1)\cos \theta^{'}_E+(n-1)\sin \theta^{'}_E ]\cos\varphi_E}\times \nonumber\\
& e^{-j\pi [(m-1)\cos \theta^{'}_B +(n-1)\sin \theta^{'}_B] \cos\varphi_B}.
\end{align}
Ideally, $\mathrm{SINR}_E$ degrades to $0$ as $\mathbf{h}^H_E \mathbf{h}_B=0$, which indicates the maximum value of the SR is reached in this case.
Toward this direction, we aware from \eqref{hehb} that the optimality condition of \eqref{opt1} achieves as the UAV deployment impels that $\textbf{h}_E^H$ is orthogonal to $\textbf{h}_B$.

However, it can be showed from (\ref{hehb}) that $\theta^{'}_E$ and $\varphi_E$ are highly coupled, leading to a tricky task in terms of simultaneously obtaining their optimal solutions.
To mitigate the difficulty of addressing the problem with respect to $\theta^{'}_E$ or $\varphi_E$, we do not jointly optimize them but solve it by an ingenious way, which is also able to obtain the optimal solutions to both $\theta^{'}_E$ or $\varphi_E$.
To guarantee the ultimate result that $\mathbf{h}^H_E \mathbf{h}_B=0$ and decouple the indivisible relationship between $\theta^{'}_E$ and $\varphi_E$, without loss of generality, we can define $\varphi_B=\varphi_E$ or $\theta^{'}_B=\theta^{'}_E$ to ease the conceived problem. For $\varphi_B=\varphi_E$ as our instance, the potential null points of array pattern along the azimuth angle dimension satisfy the following condition\cite{Kraus2002Antennas}
\begin{align}\label{aa}
\mathbf{h}^H_E \mathbf{h}_B&=\frac{1}{MN}\sum \limits_{m}^{M}e^{j\pi (m-1)(\cos \theta^{'}_E-\cos \theta^{'}_B)\cos\varphi_E}\nonumber\\
&\times\sum \limits_{n}^{N}e^{j\pi (n-1)(\sin \theta^{'}_E-\sin\theta^{'}_B)\cos\varphi_E},\nonumber\\
&=\frac{1}{MN}\cdot\frac{e^{jM\pi(\cos\theta^{'}_E-\cos\theta^{'}_B)\cos\varphi_E}-1}{e^{j\pi(\cos\theta^{'}_E-\cos\theta^{'}_B)\cos\varphi_E}-1}\nonumber\\
&\times\frac{e^{jN\pi(\sin\theta^{'}_E-\sin\theta^{'}_B)\cos\varphi_E}-1}{e^{j\pi(\sin\theta^{'}_E-\sin\theta^{'}_B)\cos\varphi_E}-1}.
\end{align}
To compel that $\mathbf{h}^H_E \mathbf{h}_B=0$, the following condition has to be satisfied, given by
\begin{align}\label{solution1}
&M\pi(\cos\theta^{'}_E-\cos\theta^{'}_B)\cos\varphi_E=\pm2k\pi,
\end{align}
or
\begin{align}\label{solutino2}
N\pi(\sin\theta^{'}_E-\sin\theta^{'}_B)\cos\varphi_E=\pm2k\pi,
\end{align}
where $k$ has to ensure that $k \neq k'M$ and $k \neq k'N$, herein $k'$ is an integer (i.e., $k' \in \mathbb{Z}$). Then, the relationship between the optimal $\theta_E'$ and the optimal $\theta_B'$ can be obtained as
\begin{align}\label{re1}
\cos\theta^{'}_E-\cos\theta^{'}_B=\frac{\pm 2k }{M \cos\varphi_E},
\end{align}
or
\begin{align}\label{re2}
\sin\theta^{'}_E-\sin\theta^{'}_B=\frac{\pm 2k}{N \cos\varphi_E}.
\end{align}
Taking $\varphi_B=\varphi_E$ and UAV flies at a constant height into account, we are aware that the X-coordinate of UAV satisfies $X_A=X_E/2$ and $\theta_B=\pi-\theta_E$, which can be readily verified according to our system model. Substituting $\theta_B=\pi-\theta_E$, $\theta^{'}_B=\theta_B-\theta_A$ and $\theta^{'}_E=\theta_E-\theta_A$ into (\ref{re1}), we obtain a correspondingly modified expression reagarding $\theta_B$, shown as
 \begin{align}\label{re5}
\cos\theta_B=\frac{\pm k }{M \cos\varphi_E \cos{\theta_A}}.
\end{align}
Alternatively, the other modified expression with respect to $\theta_B$ according to (\ref{re2}) is
\begin{align}\label{re6}
\cos\theta_B=\frac{\pm k }{N \cos\varphi_E \sin\theta_A}.
\end{align}

\emph{\textbf{Remark 1:}} Considering that $(\cos\theta^{'}_E-\cos\theta^{'}_B)\in [-2,2]$ and $\cos \varphi_E \in [0,1]$, thus $(\cos\theta^{'}_E-\cos\theta^{'}_B)\cos\varphi_E \in [-2,2]$. When $(\cos\theta^{'}_E-\cos\theta^{'}_B)\neq0$, the denominator of the first term in \eqref{aa} regarding $\theta^{'}_B$ is unequal to $0$. Naturally, $(\sin\theta^{'}_E-\sin\theta^{'}_B)\neq0$ holds. In a nutshell, our analysis for the two denominators of \eqref{aa} derives that $\sin\theta_E^{'}-\sin\theta_B^{'}\neq0$ and $\cos\theta_E^{'}-\cos\theta_B^{'}\neq 0$. Taking into account the relationship among $ \theta_A, \theta^{'}_B, \theta_B, \theta^{'}_E$ and $\theta_E$, the flight angle of UAV follows $\cos \theta_B \cos \theta_A \neq 0$ and $\cos \theta_B \sin \theta_A \neq 0$, which further arrives $\theta_A \neq \frac{p\pi}{2},~ (p\in \mathbb{Z})$, thus the denominators of \eqref{aa} are unequal to $0$. As a result, our analysis and derivation for the solution $\theta_B$ to expression \eqref{aa} is of physical significance.

Since $\varphi_E\in [0,\pi/2]$ increases as $Y_A$ increases in the domain of $(-\infty,0)$, hence $\cos\varphi_E$ decreases as $Y_A \in (-\infty,0)$ increases. Similarly, $\cos\varphi_E$ increases as $Y_A \in (0,+\infty)$ increases. With those in mind, we then have $\frac{1}{N \cos\varphi_E} \in ( 1/{N}, 1/(N \cos\varphi_E^\ast) ]$, where $\varphi_E^\ast$ refers to the optimal pitch angle for maximizing $\frac{1}{N \cos\varphi_E}$. As a matter of fact, $\varphi_E^\ast$ arrives in the case of $Y_A=0$, which further derives that
\begin{align}\label{theta0}
\cos \varphi_E^\ast= \frac{X_E/2}{\sqrt{(X_E/2)^2+g^2}}.
\end{align}
Furthermore, we note that the term $\frac{\pm k }{\sin\theta_A}$ of \eqref{re6} follows $|\frac{\pm k }{\sin\theta_A}| \in [ 1, +\infty)$, herein the left extremum arrives as $k=1$ and $\sin\theta_A =1$. Hence, we conclude that $\theta_B$ has at least one solution when  $|1/(N \cos\varphi_E^\ast)|\leq 1$.

With the above conclusion, we further derive the optimal solution $\theta_B^\ast$ to characterize $Y_A$, where $Y_A$ can be determined once $\theta_B^\ast$ is optimized.
Upon substituting the original definition $\cos\theta_B=\frac{X_A}{\sqrt{X_A^2+Y_A^2}}$ and $\cos\varphi_E=\frac{\sqrt{(X_A-X_E)^2+(Y_A)^2}}{\sqrt{(X_A-X_E)^2+(Y_A)^2+g^2}}$ into
the derivation of \eqref{re5}, we have
\begin{align}\label{re7}
\frac{X_A}{\sqrt{X_A^2+Y_A^2}} \frac{\sqrt{(X_A-X_E)^2+(Y_A)^2}}{\sqrt{(X_A-X_E)^2+(Y_A)^2+g^2}}=\frac{\pm k }{M  \cos{\theta_A}}.
\end{align}
Taking into account that $X_A=X_E/2$, $Y_A$ can be obtained as
 \begin{align}\label{re8}
Y_A=\pm \sqrt{\frac{M^2\cos^2 \theta_A X^2_E-k^2X^2_E-4k^2g^2} {4k^2} }.
\end{align}
Similarly, $Y_A$ can also be obtained in accordance with \eqref{re6} as
\begin{align}\label{re9}
Y_A=\pm \sqrt{\frac{N^2\sin^2 \theta_A X^2_E-k^2X^2_E-4k^2g^2} {4k^2} }.
\end{align}

Until now, we have obtained an analytical expression of UAV's coordinate shown of \eqref{re8} or of \eqref{re9}. It is worth noting that, in a potentially infancy stage, the right term $M^2\cos^2 \theta_A X^2_E-k^2X^2_E-4k^2g^2$ of \eqref{re8} or the term $N^2\sin^2 \theta_A X^2_E-k^2X^2_E-4k^2g^2$ of \eqref{re9} is not inherently greater than or equal to $0$, there is no solution satisfying \eqref{re8} or \eqref{re9}. However, the parameters $\theta_A$ (the yawing angle of UAV) and $g$ (the height of UAV) can be strategically regulated by UAV's attitude, hence at least one feasible solution to \eqref{re8} or \eqref{re9} is able to be gained.
Because of the analytical solution, the computational complexity is significantly reduced when compared to directly addressing problem \eqref{opt1}.
For the given $X_A$, once $Y_A$ is optimized by \eqref{re8} or \eqref{re9}, we finish the UAV deployment problem. At such an optimal coordinate $(X_A^\ast, Y_A^\ast, g)$, $\mathrm{SINR}_E$ degrades to zero while $\mathrm{SINR}_B$ achieves its maximum value $\alpha P_S/\sigma^2_B$, thus ensuring that the maximum SR performance can be achieved.

\emph{\textbf{Remark 2:}} Upon optimizing UAV deployment, we promulgate that the maximum SR performance remains achievable, even though we have not split any precious power to the AN signalling. The benefits are threefold: 1) From UAV's perspective, the hardware structure becomes more simple owing to removing the module of AN generator, which cuts down the expenditure and favors lightening the weight of UAV; 2) For the perspective of energy efficiency, all the transmit power is used to convey the confidential message, which contributes to Bob receiving a high-quality signal; 3) While for the computational complexity, our derived analytical solution to $Y_A$ on the basis of the 3-D spatial relation conspicuously mitigates the computational burden in terms of the UAV deployment.

For $\varphi_B=\varphi_E$ as our anther instance to decouple the indivisible relationship between $\theta_B=\theta_E$, the potential null points of array pattern along the pitch angle dimension satisfy the following condition ,
\begin{align}\label{pa}
\mathbf{h}^H_E \mathbf{h}_B&=\frac{1}{MN}\sum \limits_{m}^{M}e^{j\pi (m-1)\cos\theta_E(\cos \varphi_E-\cos \varphi_B)}\nonumber\\
&\cdot\sum \limits_{n}^{N}e^{j\pi (n-1)\sin\theta_E(\cos \varphi_E-\cos \varphi_B)},\nonumber\\
&=\frac{1}{MN}\cdot\frac{e^{jM\pi\cos\theta_E(\cos \varphi_E-\cos\varphi_B)}-1}{e^{j\pi\cos\theta_E(\cos \varphi_E-\cos\varphi_B)}-1}\nonumber\\
&\cdot\frac{e^{jN\pi\sin\theta_E(\cos \varphi_E-\cos \varphi_B)}-1}{e^{j\pi\sin\theta_E(\cos \varphi_E-\cos \varphi_B)}-1}.
\end{align}
Similarly, let $\mathbf{h}^H_E \mathbf{h}_B=0$, we have,
\begin{align}\label{solution1}
&M\pi\cos\theta_E(\cos \varphi_E-\cos\varphi_B)=\pm2l\pi,\nonumber\\
&l\neq l^{'}M(l^{'}=1,2,3,\ldots).
\end{align}
or
\begin{align}\label{solutino2}
&N\pi\sin\theta_E(\cos \varphi_E-\cos \varphi_B)=\pm2l\pi,\nonumber\\
&l\neq l^{'}N(l^{'}=1,2,3,\ldots).
\end{align}
Then, the relationship $\theta_E$ and $\theta_B$ can be obtained as,
\begin{align}\label{re1}
&\cos \varphi_E-\cos \varphi_B=\frac{\pm 2l }{M \cos\theta_E},\nonumber\\
&l\neq l^{'}M(l^{'}=1,2,3,\ldots).
\end{align}
or
\begin{align}\label{re2}
&\cos \varphi_E-\cos \varphi_B=\frac{\pm 2l}{N  \sin\theta_E},\nonumber\\
&l\neq l^{'}N(l^{'}=1,2,3,\ldots).
\end{align}
Note that, the azimuth angles satisfy the constraint of $\theta_B=\theta_E$. According to the system model, the Y-coordinate of Alice must satisfy $Y_A=0$. Since the azimuth angle can be adjusted with the flight angle of Alice, i.e. the flight direction which we defined as $\theta_A$, its value range is $(0,2\pi)$. The X-coordinate range of Alice is $(-\infty,0)\bigcup(X_E,+\infty)$ while $\cos \varphi_E-\cos \varphi_B$ is  a monotone decreasing function about $X_A$. Let $l=1$, it is clear that when $\frac{ 2}{M \cos\theta_E}<0$ or $\frac{ 2}{N  \sin\theta_E}<0$, the X-coordinate $X_A\in(-\infty,0)$. Conversely, when $\frac{2}{M \cos\theta_E}>0$ or $\frac{ 2}{N  \sin\theta_E}>0$, the X-coordinate $X_A\in(X_E,+\infty)$. Thus the value of $X_A$ is existed and can be easily obtained similar to the pitch angles scheme or by dichotomy method.

\section{simulation results and analysis}

In this section, we evaluate the achievable SR performance of our proposed scheme via numerical simulations, where the detailed system parameters are set as shown in Table \ref{tab}. Without loss of generality, we assume that the noise levels at Bob and Eve are identical, i.e., $\sigma^2_B=\sigma^2_E$. Moreover, the variable argument $k$ in both (\ref{re5}) and (\ref{re6}) are set to be 1.
\begin{table}[ht!]
\centering
\caption{SIMULATION PARAMETERS SETTING}\label{tab}
 \begin{tabular}{|c|c|}
\hline
Parameter&Value\\
\hline
The number of transmitter antennas ($M\times N$)&$4\times 4$ \\
\hline
Total signal bandwidth (B)&5MHz\\
\hline
Total transmit power ($P$)&1W \\
\hline
The height of UAV ($g$)&200m\\
\hline
The eavesdropper's position ($X_E,Y_E$)&(500m,0)\\
\hline
The flight angle of UAV ($\theta_A$)& $\pi/4$\\
\hline
Central carrier frequency ($f_c$)&3GHz\\
\hline
\end{tabular}
\end{table}

\begin{figure}[ht!]
\centering
\includegraphics[width=0.40\textwidth]{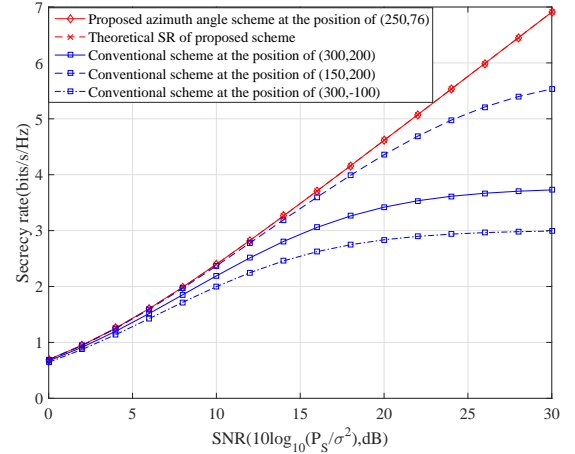}\\
\caption{The achievable SR performance versus SNR for our proposed azimuth angles scheme, where three typically random deployments are invoked as the baselines.}\label{aag}
\end{figure}
Fig. \ref{aag} shows the attainable SR performance of our proposed azimuth angles scheme versus the singal-to-noise ratio (SNR), where SNR $= 10 \log \left( P_s/\sigma^2 \right)$. For comparison, three conventionally random deployments are invoked to validate the efficiency of our proposed UAV deployment. Firstly, it can be obviously noted that our proposed UAV deployment scheme is superior to the other three random schemes in terms of the SR performance, albeit a negligible computational complexity is increased. Moreover, the SR performance gap between the proposed scheme and any other scheme becomes distinct as the SNR increases. Therefore, the main benefits of our considered UAV deployment scheme stem from not only assuring the precise transmission but also improving the security performance.
On the other hand, we note from Fig. \ref{aag} that the theoretical SR performance, i.e., the maximum performance, is coincident with that of our proposed scheme at any SNR. The result further verifies the validness of our derived solution to the UAV deployment in the context of SPWT.

\begin{figure}[ht!]
\centering
\includegraphics[width=0.40\textwidth]{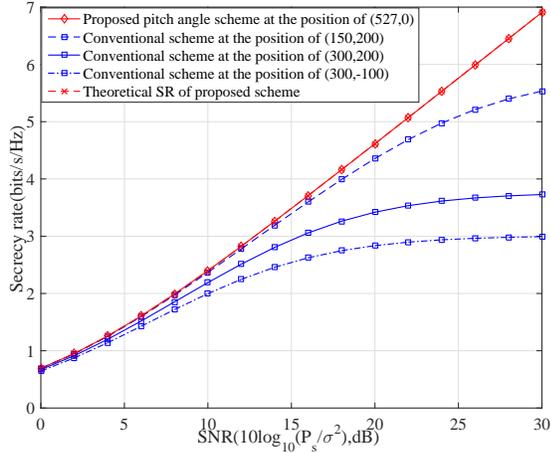}\\
\caption{The achievable SR performance versus SNR for our proposed pitch angle scheme, where three typically random deployments are invoked as the baselines.}\label{pag}
\end{figure}
Fig. \ref{pag} shows the attainable SR performance of our proposed pitch angles scheme versus the SNR. Similarly, the three conventionally random deployments are invoked for comparision. The results show that this proposed scheme is also superior to the conventionally random deployments. This also proves that the basic ideas of the two methods we proposed are completely correct. From Fig. \ref{aag} and Fig. \ref{pag}, it is clear to find, no matter the azimuth angles scheme or pitch angles scheme we adopt, if only the deployed UAV satisfies the constraint of $\mathbf{h}^H_E \mathbf{h}_B=0$ , the optimal screcy rates can achieve by the optimal beamforming of $\mathbf{v}=\mathbf{h}(\theta_B,\varphi_B)$.

\begin{figure}[ht!]
\centering
\includegraphics[width=0.40\textwidth]{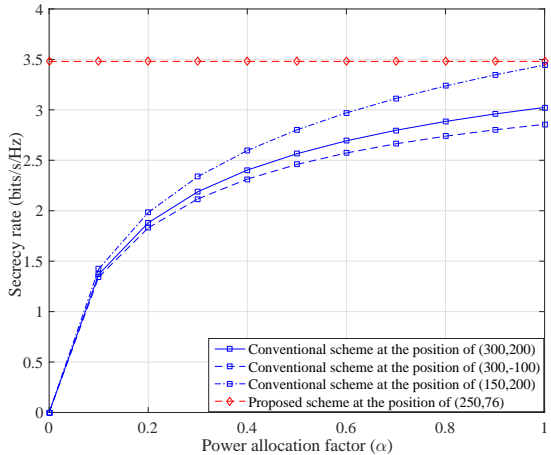}\\
\caption{The achievable SR performance versus parameter $\alpha$ for the different scheme when SNR is $15$ dB.}\label{beta1}
\end{figure}

To illustrate the efficiency of our proposed UAV-assisted SPWT scheme in gaining the security, Fig. \ref{beta1} shows that the achievable SR performance varies as $\alpha$ increases.
It can be noted that the SR performances of the three conventional comparison schemes are constantly unable to achieve the optimal SR performance even with the aid of power allocation. In fact, the SR performance of the conventional scheme seriously degrades when the UAV is randomly deployed. Hence, properly arranging UAV has a momentous impact on the SPWT. While for our proposed UAV deployment eliminating the AN signalling, it overleaps the power split operation but remains gaining the maximum SR performance, which further corroborates the potential value for the SPWT.
Another interesting conclusion is that the SR of our proposed scheme remains the maximum value and unchanged, this means our porposed scheme do not need the artificial noise, this is because $\mathbf{h}^H_E \mathbf{h}_B=0$, regardless of the power of artificial noise, the received SNR at Eve is always 0. Thus artificial noised is unnecessary, and this brings the benifit of a budget reduction.

\section{Conclusion}
In this letter, we proposed an UAV deployment scheme in the context of SPWT from the perspective of azimuth angle and pitch angle. In this scheme, we abandoned FDA for the first time and adopted the method of combining DM with 3D scenario, which reduces the system budget significantly. Compared to the conventional method, the proposed scheme is more superior in terms of the attainable SR performance. Moreover, our proposed UAV deployment algorithm gives the analytical solutions which has almost no complexity, this is also an important benifit of our proposed scheme. Interestingly, although we introduced AN in this hybrid SPWT system, the methmatic analysis shows that when the allocated power in AN is zero, the performance achieves the optimal, this means our proposed scheme do not need AN assistance to achieve the optimal SR performance, which has a powerful ability in economizing the precious energy resource.

\ifCLASSOPTIONcaptionsoff
  \newpage
\fi

\bibliographystyle{IEEEtran}
\bibliography{IEEEfull,reference}
\end{document}